\documentclass{aa} 
\usepackage{epsfig}
\usepackage{graphicx}
\usepackage{natbib}
\usepackage{amssymb}

\begin{document}

                          
\title{PSRs J0248+6021 and J2240+5832: Young pulsars in the northern Galactic plane}
\subtitle{Discovery, timing, and gamma-ray Observations}

\author{
G.~Theureau$^{(1)}$ \and 
D.~Parent$^{(2, 3, 4)}$ \and 
I.~Cognard$^{(1)}$ \and 
G.~Desvignes$^{(1,5)}$ \and 
D.~A.~Smith$^{(4)}$ \and 
J.~M.~Casandjian$^{(6)}$ \and 
C.~C.~Cheung$^{(2, 7)}$ \and 
H.~A.~Craig$^{(8)}$ \and 
D.~Donato$^{(9, 10)}$ \and 
R.~Foster$^{(11)}$ \and
L.~Guillemot$^{(4, 12)}$ \and 
A.~K.~Harding$^{(13)}$ \and
J.-F.~Lestrade$^{(14)}$ \and 
P.~S.~Ray$^{(2)}$ \and 
R.~W.~Romani$^{(8)}$ \and 
D.~J.~Thompson$^{(13)}$ \and 
W.~W.~Tian$^{(15,16)}$ \and
K.~Watters$^{(8)}$ 
}
\authorrunning{LAT collaboration}

\institute{
\inst{1}~Laboratoire de Physique et Chemie de l'Environnement et de l'Espace, LPC2E UMR 6115 CNRS, F-45071 Orl\'eans Cedex 02, and Station de radioastronomie de Nan\c{c}ay, Observatoire de Paris, CNRS/INSU, F-18330 Nan\c{c}ay, France\\ 
\inst{2}~Space Science Division, Naval Research Laboratory, Washington, DC 20375, USA\\ 
\inst{3}~George Mason University, Fairfax, VA 22030, USA\\ 
\inst{4}~Universit\'e Bordeaux 1, CNRS/IN2P3, Centre d'\'Etudes Nucl\'eaires de Bordeaux, Gradignan, CENBG, Chemin du Solarium, BP 120, 33175 Gradignan, France\\ 
\inst{5}~Department of Astronomy and Radio Astronomy Laboratory, University of California, Berkeley, CA 94720, USA\\
\inst{6}~Laboratoire AIM, CEA-IRFU/CNRS/Universit\'e Paris Diderot, Service d'Astrophysique, CEA Saclay, 91191 Gif sur Yvette, France\\  
\inst{7}~National Research Council Research Associate, National Academy of Sciences, Washington, DC 20001, USA\\ 
\inst{8}~W. W. Hansen Experimental Physics Laboratory, Kavli Institute for Particle Astrophysics and Cosmology, Department of Physics and SLAC National Accelerator Laboratory, Stanford University, Stanford, CA 94305, USA\\ 
\inst{9}~Center for Research and Exploration in Space Science and Technology (CRESST) and NASA Goddard Space Flight Center, Greenbelt, MD 20771, USA\\ 
\inst{10}~Department of Physics and Department of Astronomy, University of Maryland, College Park, MD 20742, USA\\ 
\inst{11}~High Performance Technologies, Inc., 11955 Freedom Drive, Reston, VA 20190-5673, USA\\ 
\inst{12}~Max-Planck-Institut f\"ur Radioastronomie, Auf dem H\"ugel 69, 53121 Bonn, Germany\\ 
\inst{13}~NASA Goddard Space Flight Center, Greenbelt, MD 20771, USA\\ 
\inst{14}~Observatoire de Paris-CNRS/LERMA, 77 av. Denfert Rochereau, Paris, 75014, France\\ 
\inst{15}~Department of Physics and Astronomy, University of Calgary, Calgary, AB, Canada T2N 1N4\\ 
\inst{16}~National Astronomical Observatories of China, Beijing, 100012, China\\ 
\email{smith@cenbg.in2p3.fr} \\
\email{dparent@ssd5.nrl.navy.mil} \\
\email{theureau@cnrs-orleans.fr} \\
}

\date{Received:   \hspace*{3cm}; accepted: }

\authorrunning{Theureau et al.}
\titlerunning{PSRs~J0248+6021 and J2240+5832}

\abstract
{Pulsars PSR~J0248+6021 (with a rotation period $P=217$ ms and spin-down power $\dot E = 2.13 \times 10^{35}$ erg s$^{-1}$) 
and PSR~J2240+5832 ($P=140$ ms, $\dot E = 2.12 \times 10^{35}$ erg s$^{-1}$) were discovered
in 1997 with the Nan\c cay radio telescope during a northern Galactic plane survey, 
using the Navy-Berkeley Pulsar Processor (NBPP) filter bank. 
The GeV gamma-ray pulsations from both were discovered using the \textit{Fermi} Large Area Telescope. 
 }
{We characterize the neutron star emission using radio and gamma-ray observations, 
and explore the rich environment of PSR~J0248+6021.}
{Twelve years of radio timing data, including glitches, with steadily
improved instrumentation, such as the Berkeley-Orleans-Nan\c cay pulsar backend,
and a gamma-ray data set $2.6$ times larger than previously published allow detailed
investigations of these pulsars. Radio polarization data allow comparison with the
geometry inferred from gamma-ray emission models.}
{The two pulsars resemble each other in both radio and gamma-ray data.
Both are rare in having a single gamma-ray pulse offset far from the radio peak.
The anomalously high dispersion measure for PSR~J0248+6021 (DM $ = 370$ pc cm$^{-3}$)
is most likely due to its being within the dense, giant HII region W5 in the Perseus arm at a distance
of 2 kpc, as opposed to being beyond the edge of the Galaxy as obtained from models of average electron distributions.
Its large transverse velocity and the low magnetic field along the line-of-sight favor this small distance. 
Neither gamma-ray, X-ray, nor optical data yield evidence of a pulsar wind nebula surrounding
PSR~J0248+6021. We report the discovery of gamma-ray pulsations from PSR~J2240+5832. We argue that it could be in the 
outer arm, although slightly nearer than its DM-derived distance, but that it may be in the Perseus arm
at half the distance.
}
{The energy flux and distance yield a
gamma-ray luminosity for PSR~J0248+6021 of $L_\gamma = (1.4 \pm 0.3)\times 10^{34}$ erg s$^{-1}$.
For PSR~J2240+5832, we find either 
$L_\gamma = (7.9 \pm 5.2) \times 10^{34}$ erg s$^{-1}$ if the pulsar is in the outer arm, or
$L_\gamma = (2.2 \pm 1.7) \times 10^{34}$ erg s$^{-1}$ for the Perseus arm.
These luminosities are consistent with an $L_\gamma \propto \sqrt{\dot E}$ rule.
Comparison of the gamma-ray pulse profiles with model predictions, including the
constraints obtained from radio polarization data, implies outer magnetosphere emission.
These two pulsars differ mainly by their inclination angles and acceleration gap widths, which in
turn explain the observed differences in the gamma-ray peak widths. 
}


\maketitle

\section{Introduction}
Pulsars PSR~J0248+6021 and PSR~J2240+5832 were discovered with the Nan\c cay radio telescope 
(NRT) in a survey of the northern Galactic plane from 1997 to 1999 \citep{1997AAS...19111110F}.
\citet{1999AAS...194.5206R} reported their locations and rotation periods.
Twelve years of timing measurements allow us to provide accurate ephemerides, including a 
proper motion determination for PSR~J0248+6021 that helps us understand its relation to the 
complex of gas clouds co-located on the sky. Polarization data provides geometry constraints
in addition to those obtained by comparing the gamma-ray pulse profiles with model predictions.

By coincidence, both pulsars have a high spin-down power $\dot E = 2.1 \times 10^{35}$ erg s$^{-1}$,
making it also likely that they emit in GeV gamma-rays, as discussed by \citet{DAS08}.
Indeed, PSR~J0248+6021 is among the 46 gamma-ray pulsars described in the ``First Pulsar Catalog'' \citep{FermiPsrCata}
using the Large Area Telescope (LAT) on the \textit{Fermi} Gamma-ray Space Telescope (formerly GLAST). 
They have the same magnetic field strengths at the neutron-star
light cylinder, $B_{LC}$, to within 20\%, and the characteristic ages $\tau_c = P/2\dot P$ differ by only a factor of 
two. Table \ref{Table} lists measured and derived parameters for both pulsars.
PSR~J2240+5832 has a smaller dispersion measure (DM $ = 263.5$ pc cm$^{-3}$) than 
PSR~J0248+6021 (DM $ = 370$ pc cm$^{-3}$), and the NE2001 model of the
Galactic electron distribution \citep{Cordes2002} assigns PSR~J2240+5832 a distance of $10.3_{-3.3}^{+\infty}$ kpc,
whereas it places PSR~J0248+6021 well beyond the edge of the Galaxy ($>43.5$ kpc).
Curiously however, it is the nominally closer pulsar, 
PSR~J2240+5832 that pulsates \textit{less} brightly in the \textit{Fermi} data.

Two obvious solutions to this seeming paradox are that either there is some intrinsic difference between
the pulsars, such as their orientation angles and beaming, or that the distances are simply misunderstood. 
In this paper, we clarify the situation.
\section{Nan\c cay northern Galactic plane survey}
%
%
The NRT is a meridian instrument, with a primary antenna 200 m wide by 35 m tall, 
equivalent to a parabolic dish with a 94 m diameter.
It can track objects with declinations $\delta > -39^\circ$ for roughly one hour around culmination.
The half-power beam width at 1.4 GHz is 4 arcmin (east-west) by 22 arcmin (north-south) at 
$\delta = 0^\circ$, a shape well-adapted to sky scanning. 
The 1.4 GHz nominal system temperature at the time of 
these observations was about 50 K in both horizontal and vertical polarizations,
for $\delta \simeq 60^\circ$. 
The ``FORT'' receiver upgrade from 1999 to 2001 improved sensitivity by a factor
of $2.2$,  with an efficiency of $1.4$ K/Jy and a system temperature of 35 K 
at $1.4$ GHz. 
The frequency coverage is now continuous from $1.1$ GHz to 3.5 GHz and allows full Stokes measurements
\citep{2005A&A...430..373T}.

Two pulsar instrumentation systems (``backends'') were used successively, 
first for the survey and then for the continued timing.
The earlier system was the Navy-Berkeley Pulsar Processor (NBPP) filter bank,
designed and built for this survey at the Naval Research Laboratory (NRL) in collaboration with the
University of California, Berkeley.
We set the 96 channels in the full crate to cover $1.5$ MHz/channel, for a total bandwidth of 144 MHz 
\citep{1997PASP..109...61B}.
The output data has a time resolution of 50 to 100 $\mu$s in search mode, allowing
detection of signals with periods as short as 0.1 ms.
For the survey, data were acquired at a sampling rate of 60 $\mu$s using 4-bit digitization of
96 channels and both polarizations.

Since 2004, the NRT pulsar backend is the BON (Berkeley-Orl\'eans-Nan\c cay) 
coherent dedispersor with a bandwidth of 128 MHz. A spectrometer 
digitizes data voltages, followed by four data servers that share the data over a 
70-node cluster of personal computers (PCs) running the Linux operating system. 
In 2009, a GPU-based (Graphics Processor Unit) computer system replaced the PC cluster, 
with the same computing power in only two nodes.
Dedispersion is performed with a special filter in the complex Fourier domain. 
The timing resolution is a few 100 ns on the most stable 
millisecond pulsars \citep{2006IAUJD...2E..36C,CognardBlois2009}.

\begin{figure}[htbp]
\begin{center}
\epsfig{file=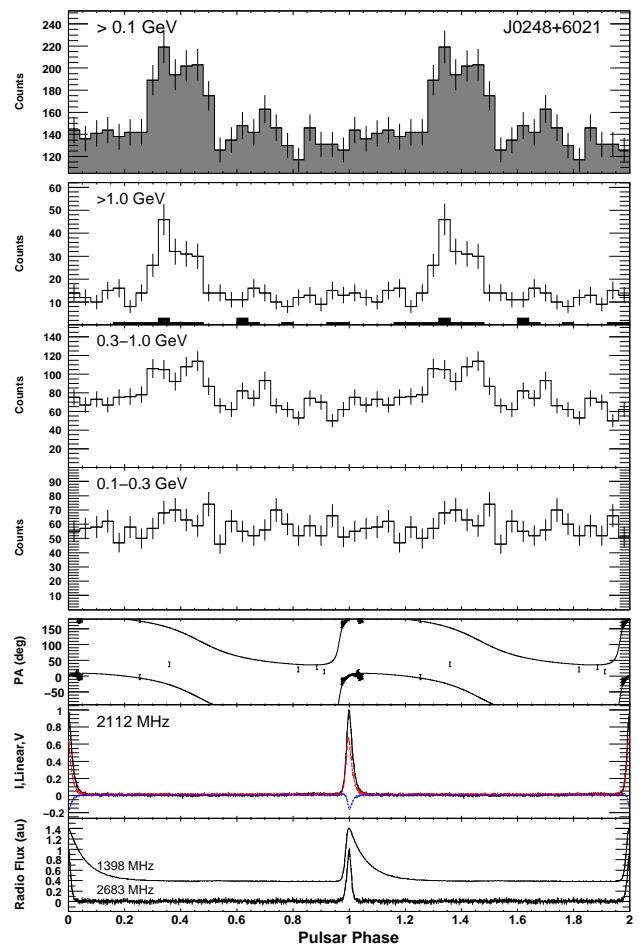,width=9.0cm}
\caption{Phase-aligned $\gamma$-ray and radio light curves for PSR J0248+6021 obtained with 
the \emph{Fermi} Large Area Telescope and the Nan\c cay Radio Telescope. 
The bottom panels show the radio profiles at three frequencies used to build the ephemeris. The second panel from the bottom
shows the degree of linear (red dashed) and circular
polarizations (blue dotted), as well as the linear polarization position angle and a Rotating Vector Model (RVM) fit.
The other panels show the phase-folded
$\gamma$-ray data in different energy bands. Two rotations are shown for clarity.}
\label{phasos_0248}
\end{center}
\end{figure}

%
\begin{figure}[htbp]
\begin{center}
\epsfig{file=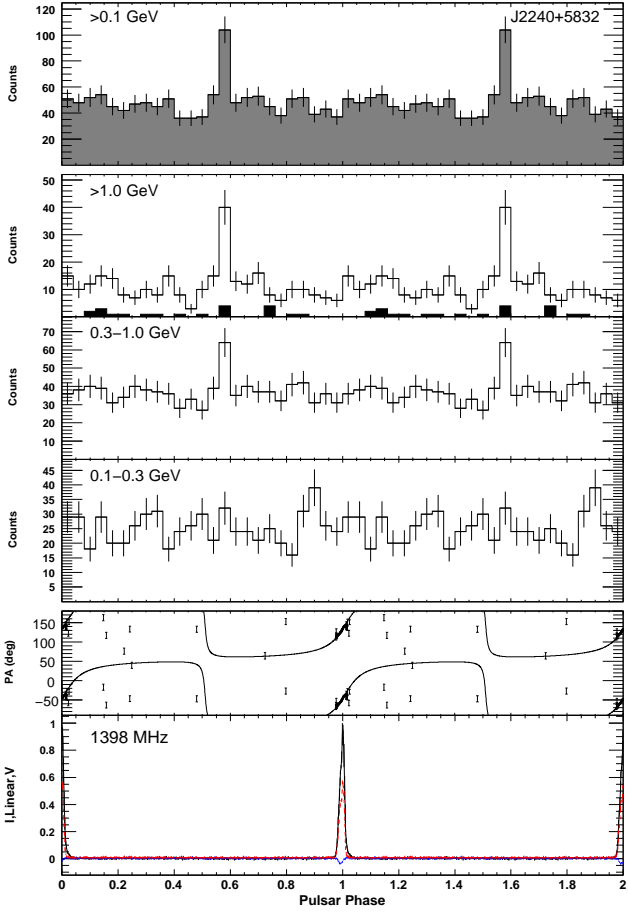,width=9cm}
\caption{Phase-aligned light curves for PSR J2240+5832. 
The bottom panel shows the 1.4 GHz radio profile, 
along with the linear (red dashed) and circular polarizations (blue dotted).
The second panel from the bottom shows the linear polarization position angle sweep with an RVM fit. 
The other panels show the phase-folded $\gamma$-ray data obtained with the \emph{Fermi} Large Area Telescope in 
different energy bands. Two rotations are shown.}
\label{phasos_2240}
\end{center}
\end{figure}
The NBPP was used to conduct a high-frequency, wide-bandwidth (1280--1430 MHz) pulsar survey 
of the Galactic plane from 1996 to 1998 \citep{1996ASPC..105...25F}. 
The search covered longitudes  $-15^\circ < l < 180^\circ$ for Galactic latitudes $|b| < 3^\circ$, 
for a total sky coverage of 1170 square degrees. The survey was optimized to find
distant pulsars with periods as short as the theoretical 
break-up speed of a neutron star. 
Over 40 000 pointings were acquired, each with an integration time of 2 minutes, 
for an expected $5\sigma$ detection of a $0.5$ mJy peak intensity profile after dedispersion. 
Observations were completed in late 1998. The final data volume is $\sim 5$ Tbytes. 
\begin{figure}[htbp]
\begin{center}
\epsfig{file=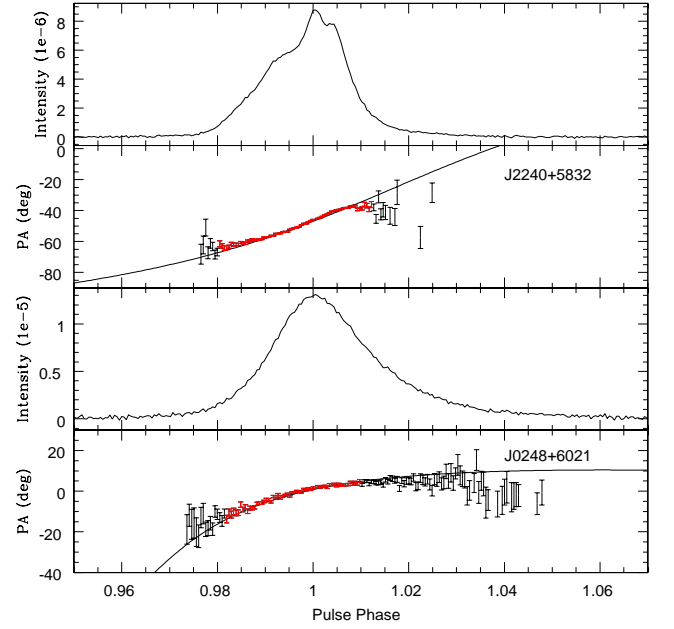,width=9cm}
\caption{Expanded view of the radio polarization position angle sweep near the peak in radio intensity. 
The red points show the data used in the RVM fit. The black points failed the selection cuts
described in the text. Top two frames: For PSR J2240+5832, at 1.4 GHz. The RVM curve shown corresponds to
inclination angles $\alpha = 108^\circ$ and $\zeta = 123^\circ$. Bottom two frames: For PSR J0248+6021,
at 2.1 GHz. The RVM curve shown corresponds to inclination angles $\alpha = 46^\circ$ and $\zeta = 52^\circ$.
}
\label{PolarZoom}
\end{center}
\end{figure}
\begin{center}
\begin{figure}
\epsfig{file=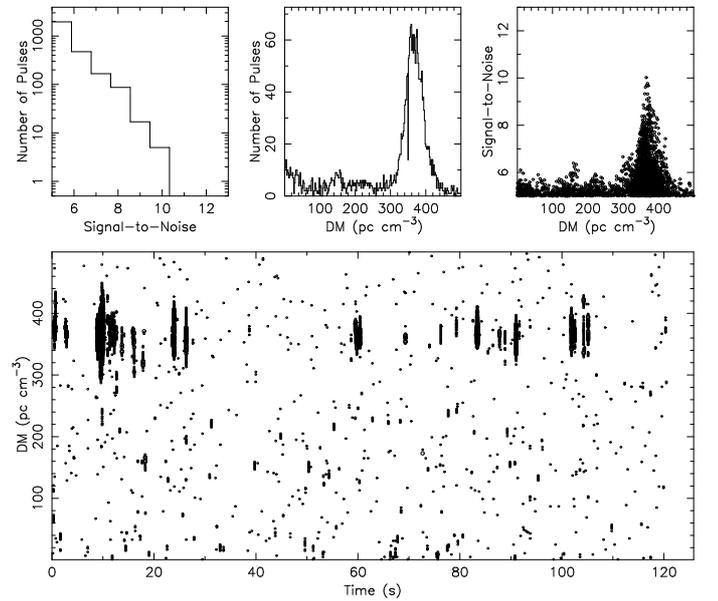,width=9cm}
\caption[]{A single pulse search of PSR~J0248+6021.
For each of 250 DM values, the 96 NBPP radio frequency samples are de-dispersed and summed. Bottom frame:
Summed intensities with signal-to-noise $5\sigma$ above the average noise level are shown as points.  
Larger points indicate larger S/N.
Top frames: Projections showing a clear pulsar detection.}
\label{singlepulses}
\end{figure}
\end{center}
%
A total of two pulsars were discovered in the survey, during the first 1998 data analysis based
on a fast Fourier transform (FFT).
Figs. \ref{phasos_0248} and \ref{phasos_2240} show multiwavelength pulse profiles 
for PSRs J0248+6021 and J2240+5832, respectively. Table \ref{Table} lists various properties.
PSR J0248+6021 has a spin period of 217 ms, with a duty cycle of 12\% at 1.4 GHz.
A scattering tail is seen at low frequency, as expected from the large DM and confirmed by the observations
made at higher frequencies. PSR J2240+5832 has a spin period of 140 ms and the radio peak is even narrower, 7\% of a rotation at half-maximum. 
Relatively few pulsars are known in this direction ($l = 106.6^\circ$) and none with such a large DM.
It may be one of the few pulsars known in the outer arm of the Galaxy and amongst the most distant known
gamma-ray pulsars.

Subsequent FFT re-analysis using the \texttt{PRESTO} package 
developed by \cite{2003ApJ...589..911R} identified no additional pulsars.
We also used PRESTO to search for single pulses from the entire dataset:
raw data were first dedispersed for 250 values ranging from 0 to 500 pc cm$^{-3}$ keeping 
the original 60 $\mu$s resolution. Each time sequence was then searched for single 
pulses with a signal-to-noise ratio (hereafter S/N) greater than 5. The search used matched filtering
with boxcars up to 30 bins allowing good sensitivity up to 1.8 ms pulse width.
We clearly detected series of very strong bursts from PSR J0248+6021 at the pulsar DM as shown in Fig. \ref{singlepulses}.
No other single pulse signal was seen, including from PSR J2240+5832.
PSR J0248+6021 is thus somewhat rare, in that it seems to
burst in a way similar to PSR~B0656+14 and RRATs \citep[Rotating Radio Transients,][]{2006ApJ...645L.149W}.
The single pulse search is confirmed to be complementary to a classical FFT analysis,
favoring pulsars with a low average signal strength but having strong bursts.

\begin{figure}[t]
  \centering
  \includegraphics[width=9cm,angle=270]{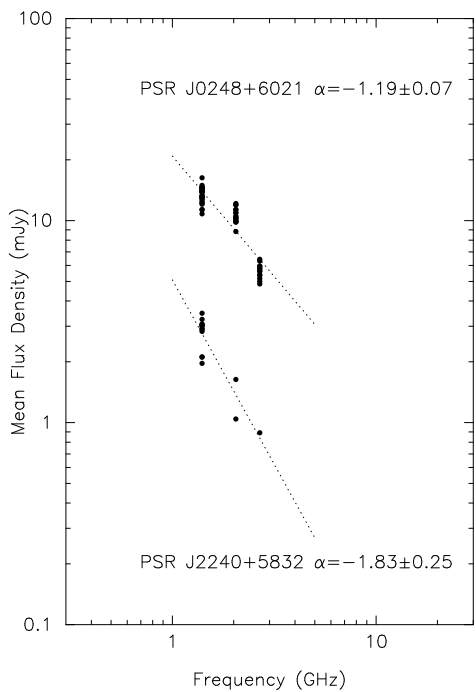}
  \caption{Calibrated mean radio flux densities as a function of observation frequency. 
Note that the sign convention for the index is the opposite of that used for the gamma-ray spectra.}
\label{0248-2240-flx}
\end{figure}
  

\section{Radio measurements}
\subsection{Flux density}
The radio flux for pulsar profiles obtained with the Nan\c cay coherent dedispersion instrumentation
can be accurately measured using a pulsed noise diode.
The diode's counts-to-mJy scaling is obtained from fiducial galaxies.
The diode is fired at 3.3 Hz for 15 seconds at the start of each observation,
then phase-folded as for a pulsar.
Several dozen observations at 1.4, 2.05, and 2.68 GHz were selected for PSR J0248+6021.
The 1.4 GHz average value for the mean ({\em i.e.}, phase-averaged) flux density of $13.7\pm 2.7$ mJy
is shown in Fig. \ref{0248-2240-flx}, and a spectral index of $-1.19 \pm 0.07$ is obtained.
These values are consistent with those obtained previously 
using the radiometer equation for both polarization channels, along with the known telescope characteristics 
($T_{sys} = 20.5$ Jy) and an assumed continuum flux of 6.5 Jy at 1.4 GHz in this sky direction 
\citep{1982A&AS...48..219R}, to convert rms noise fluctuations in the off-pulse part 
of the profile to a Jansky scale. 
 
Fourteen observations for PSR J2240+5832 yield a 1.4 GHz average value
for the mean flux density of $2.7 \pm 0.7$ mJy, and a spectral index of $-1.83 \pm 0.25$, as
also shown in Fig. \ref{0248-2240-flx}. The sign convention for the radio energy spectral index is
the opposite of that used for the gamma-ray photon index (i.e., $S(\nu) \propto \nu^\alpha$ as compared to $E^{-\Gamma}$
as in Equation 2 in Section 5).

\begin{figure*}
\centerline{  \includegraphics[width=13.0cm]{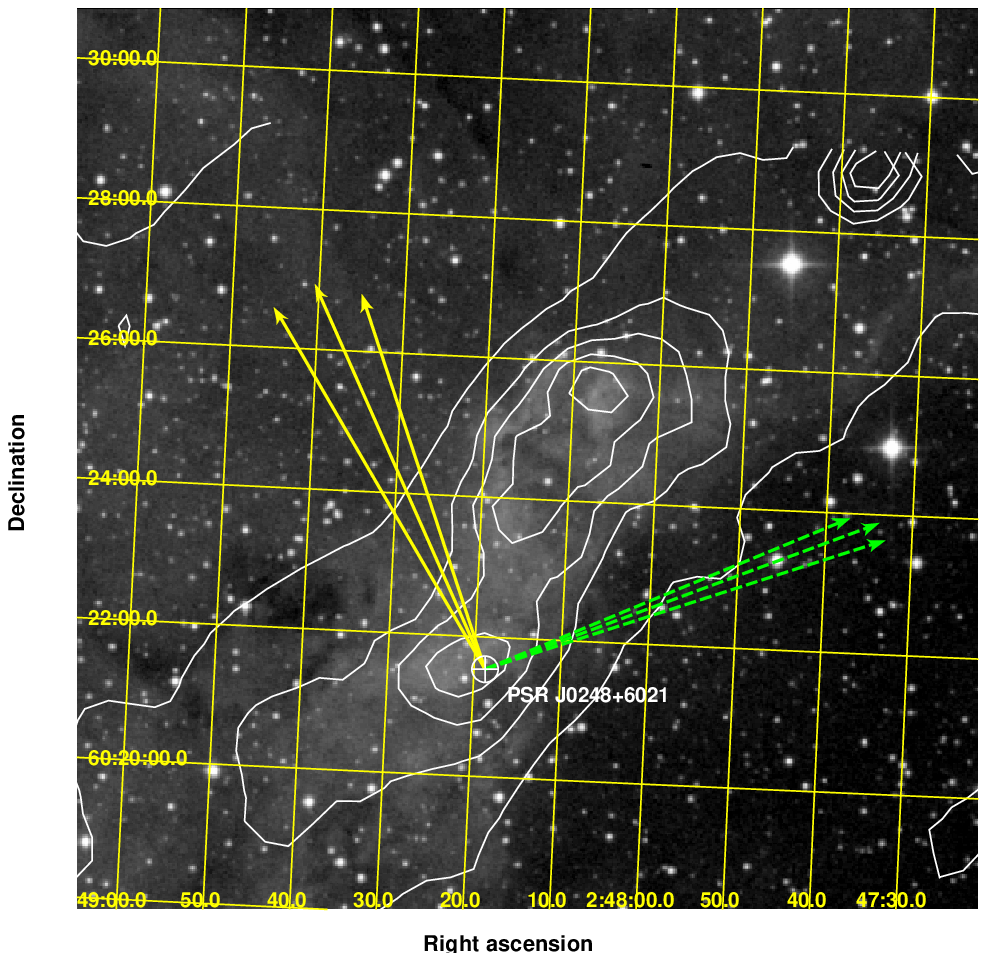}
  \includegraphics[width=6.0cm]{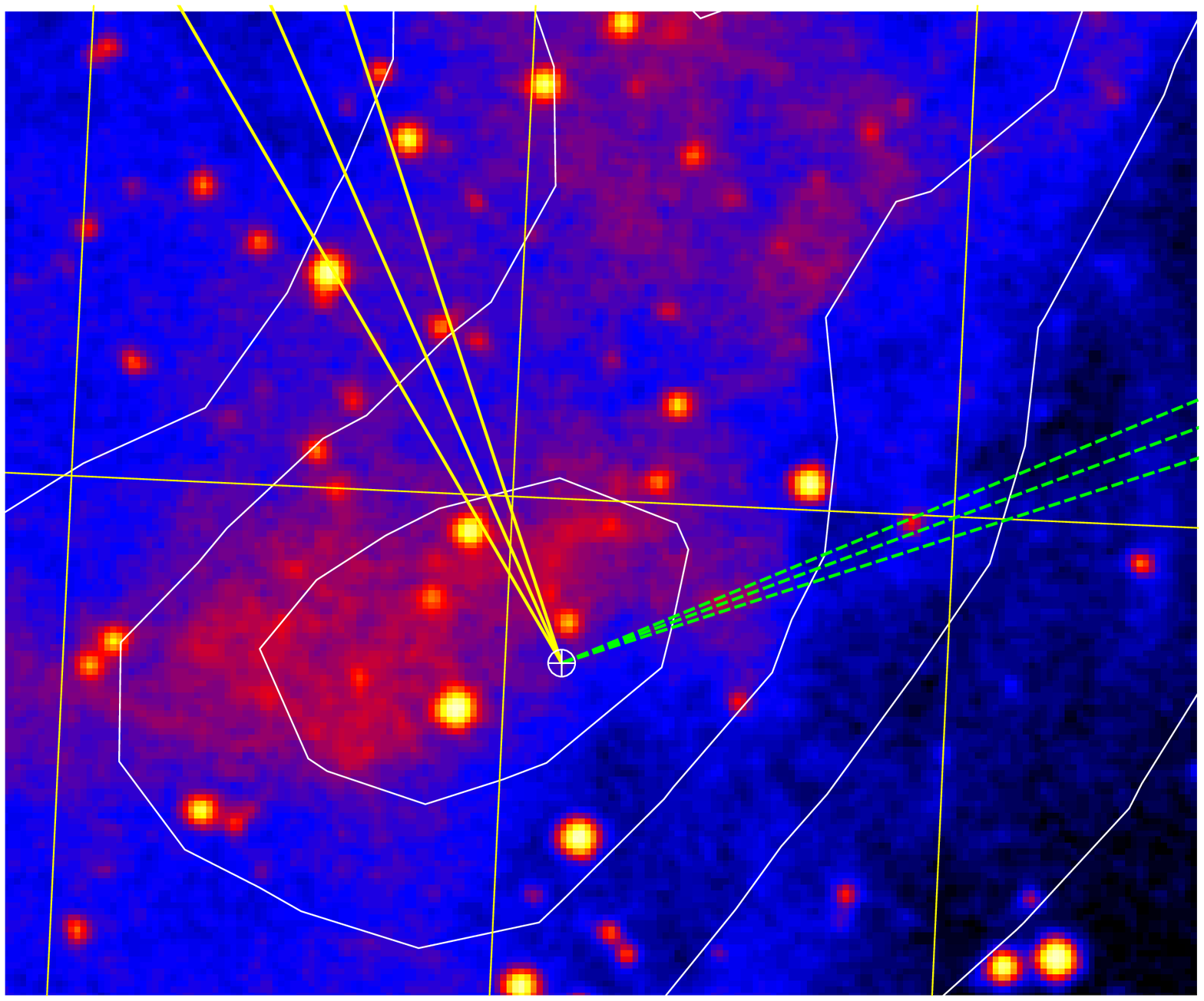} }
\caption{
Environment of PSR J0248+6021 from red Palomar Optical Sky Survey photographic plates (POSS-II, background) and 
NRAO VLA Sky Survey (NVSS) 1.4 GHz charts (contours).
The solid arrows show the proper-motion position angle and uncertainty. The dashed arrows show the polarization
position angle at the inflection point and uncertainty, extrapolated to infinite frequency. The arrow lengths have no meaning. The zoomed
image at right shows the star near the pulsar excluded as an optical counterpart, as well as the optical arc
emanating from the star discussed in the text.}
\label{FigW5}
\end{figure*}

\subsection{Polarization}
Fig. \ref{PolarZoom} zooms in on the sweep with phase of the linear polarization position angle (PA)
for both pulsars. PA data is shown for all points with a minimum S/N
in Figs. \ref{phasos_0248} and \ref{phasos_2240}.
We reduced the data using PSRCHIVE \citep{PSRCHIVE}. 
For PSR J0248+6021, we averaged 399 minutes of data from MJD 54851 to 55182, and 
obtained a rotation measure of $RM=-158 \pm 6$ rad m$^{-2}$ ($3\sigma$ statistical uncertainties). 
At inflection, extrapolated to infinite frequency, we obtain PA$_0 = -67^\circ \pm 3^\circ$,
shown in Fig. \ref{FigW5} together with the proper motion direction, discussed below.
For PSR J2240+5832, we used 846 minutes of 1.4 GHz data recorded from MJD 54866 to 55209,
yielding $RM=24 \pm 4$ rad m$^{-2}$ and PA$_0 = -39^\circ \pm 2^\circ$.

We fit the PA and phase data with the rotating vector model \citep[RVM,][]{RVM} to estimate the inclination $\zeta$
of the neutron-star rotation axis to the line-of-sight to Earth, and the angle $\alpha$ between the rotation
axis and the magnetic axis. We use only data points with a minimum S/N of $\sigma_{\rm PA} < 2^\circ$. 
For PSR J0248+6021, the PA sweep inflection point seems to lead the radio peak
due to bias induced by the broad scattering tail prominent at $1.4$ GHz. 
We therefore use the $2.1$ GHz data for the RVM analysis.
The fit results still vary with the choice of data in the tail but become stable
when we favor the leading edge of the PA swing. We interpret this as a deformation due to
residual scattering. We reject the remaining points with phase $0.01$ beyond the peak.
The results are robust once the choice to cut the tail is made.
The points used for the RVM fit are shown in red in Fig. \ref{PolarZoom}.
As usual for RVM fits, the magnetic impact parameter $\beta=\zeta-\alpha$ is most strongly constrained,
here to $\beta \approx +5^\circ$. A wide range of $\alpha$ from $40^\circ$ to $80^\circ$ provides comparably good fits, 
and acceptable fits extend from $\alpha \approx 25^\circ$ to $110^\circ$. The best fit has
$\chi^2 = 86.6$ (reduced $\chi^2=1.6$).

The narrow radio pulse of PSR J2240+5832 is less affected
by scattering: a fit to the 1.4 GHz profile yields $\beta = 16^\circ$ (for typical $\alpha$). 
Dispersion in the trailing data points causes a poor fit, with a minimum at $\chi^2=329$ (reduced $\chi^2=5.1$).
Comparable fits persist from $\alpha \approx 75^\circ$ to $130^\circ$, and plausible
solutions extend from $\alpha\approx 10^\circ$ to $150^\circ$.'

\begin{figure*}[htbp]
\begin{center}
\epsfig{file=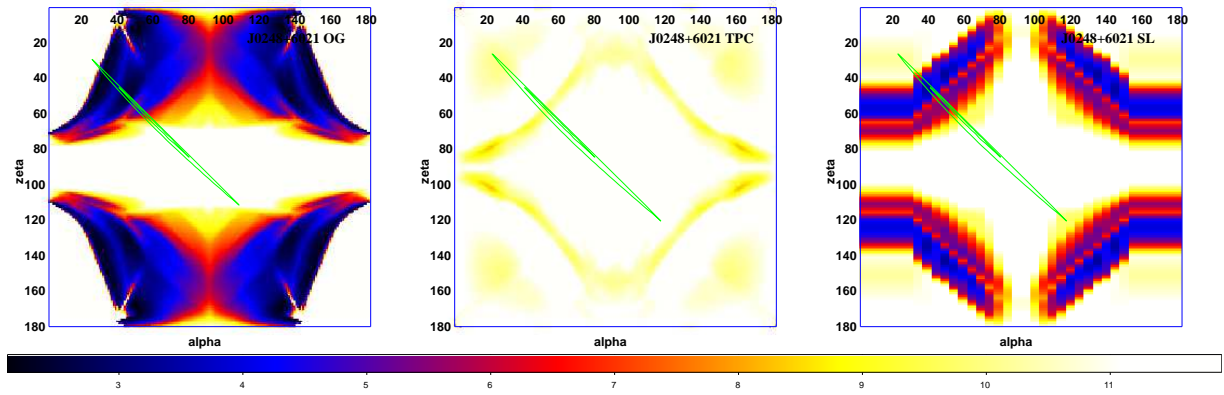,width=16.0cm}
\caption{
Pulsar geometry and emission modeling for PSR J0248+6021. Green 
contours show the rotating vector model fit to the radio polarization data 
(contours at $\delta(\chi^2/{\rm DoF}) = +0.25,\, +0.5$ above the minimum $\chi^2/{\rm DoF}$ of 1.6).
The color backgrounds are $\chi_3$ maps of the fit to the observed $>100$\,MeV
pulse profile to the outer gap model (left),
the two-pole caustic model (middle), and the separatrix layer model (right), 
for different values of the magnetic inclination,
$\alpha$, and the minimum angle to the line-of-sight, $\zeta$ \citep{AtlasII}.
Each panel has the same color scale, where dark colors represent better fits. The
preferred models lie along the green RVM-selected band.
}
\label{J0248_OGTPC}
\end{center}
\end{figure*}

\begin{figure*}[htbp]
\begin{center}
\epsfig{file=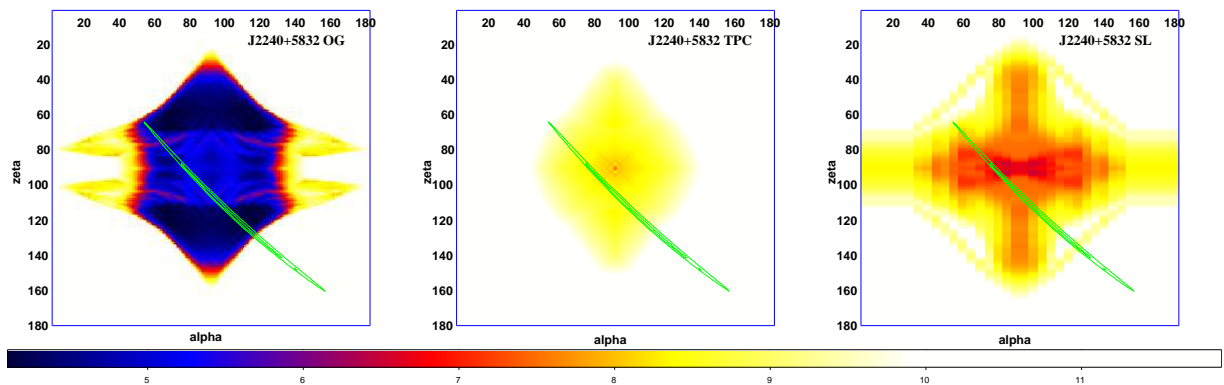,width=16.0cm}
\caption{
Pulsar geometry and emission modeling for PSR J2240+5832, as in Fig. \ref{J0248_OGTPC}. The green
contours of the RVM fits lie $\delta(\chi^2/{\rm DoF}) = +0.4,\, +0.8$ above 
the minimum of $\chi^2/{\rm DoF}=5.1$.
}
\label{J2240_OGTPC}
\end{center}
\end{figure*}

The green contours in Figs. \ref{J0248_OGTPC} and \ref{J2240_OGTPC} summarize
the RVM fit results, while the RVM curves in Fig. \ref{PolarZoom} show
the fit for specific $(\alpha, \zeta)$ combinations within the favored region.
These classic RVM fits assume radio emission at low altitudes. It is becoming increasingly
clear \citep{2007MNRAS.380.1678K,2010ApJ...716L..85R} that
for many young pulsars, especially the gamma-ray detected pulsars,
the radio emission arises at a substantial fraction of the light cylinder radius, $R_{LC}$.
This in turn means that the polarization PA sweeps are offset significantly
from the radio pulse \citep{1991ApJ...370..643B},
and that more subtle distortions in the PA sweep shape may occur.
Self-consistent fitting using large radio altitudes should therefore
lead to a shift in the inferred $\alpha$ and $\zeta$ values.

\subsection{Radio timing and proper motions}
\begin{center}
\begin{figure}
\epsfig{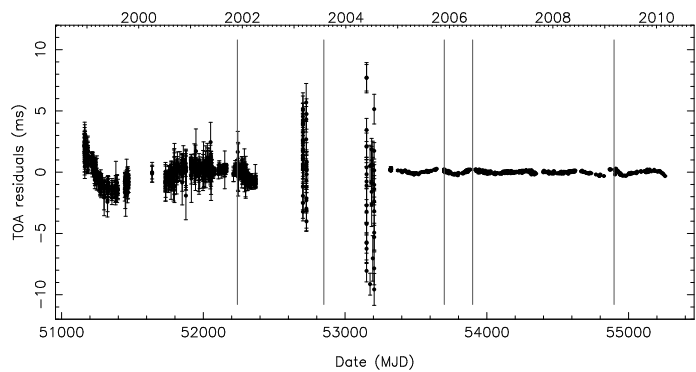}
\caption[]
{Time-of-arrival post-fit residuals for PSR J0248+6021 when glitch parameters are fit.
Vertical lines indicate glitch epochs.
The timing resolution of the datasets using the NBPP (left), BON prototype (middle), and BON (right)
backends is apparent.}
\label{postfit}
\end{figure}
\end{center}
\begin{center}
\begin{figure}
\epsfig{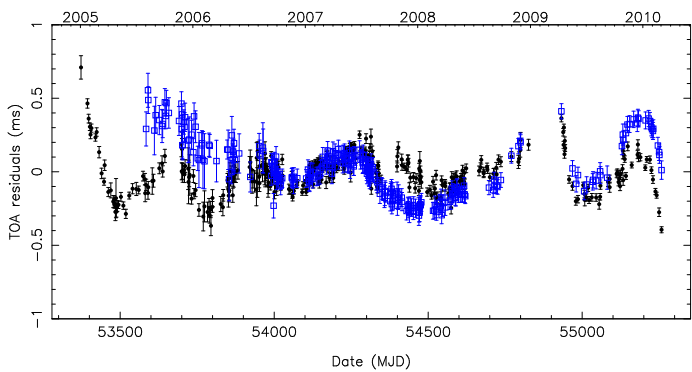}
\caption[]
{Time-of-arrival post-fit residuals for PSR J0248+6021, allowing for linear variations of dispersion measure over time, 
as in Fig. \ref{postfit}, but showing only data acquired with the BON backend. 
The $2.0$ GHz data (blue squares) are clearly above the $1.4$ GHz data  (black dots) for
the year 2006 and for 2009-2010, but below these data in 2008. We interpret the increasing, then decreasing DM value
as evidence that the neutron star is moving through a dense environment.
}
\label{ZoomedResiduals}
\end{figure}
\end{center}

Both pulsars were observed regularly for twelve years with increasingly advanced instrumentation, 
except for 18 months in 2003-2004 when only the BON prototype with a 16 MHz bandwidth was available. 
Fig. \ref{postfit} shows the evolution of the timing precision over the years, and also indicates
five principal glitch epochs for PSR J0248+2240. 
The largest was $\delta\nu / \nu = 7.5\times 10^{-7}$ in March 2009 (MJD 54897.41), where $\nu$ is the rotation
frequency and $\delta\nu$ is the permanent pulse frequency increment. The epochs and $\delta\nu / \nu$ values for the other
four glitches were 
$(52240.0, \, 1.6\times 10^{-10}), \, (52850.0, \,  8.2\times 10^{-9}), \, (53700.0, \,  6.9\times 10^{-11})$ and
$(53900.0, \,  -2.6\times 10^{-11})$.
Fig. \ref{ZoomedResiduals} shows the BON data, including the data used to phase-fold the gamma-ray photons, in greater detail.
The timing model includes a first time derivative of the DM (``DM1'' variable in TEMPO2).
Higher-order DM variations as well as achromatic timing noise can be seen.
For PSR J2240+5832, no large glitches were seen, but even after whitening some residual structure due to timing noise remains.

\begin{figure}[t]
  \centering
  \includegraphics[width=5.5cm,angle=270]{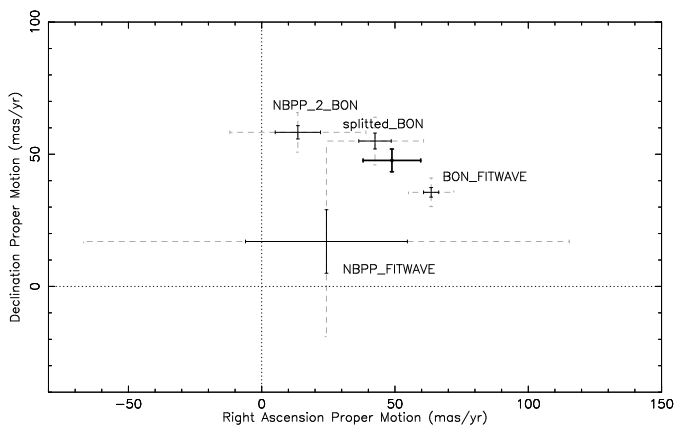}
  \caption{The PSR~J0248+6021 proper motion measured by different methods (Section 3.3). 
The weighted average of $\mu = \sqrt{\mu_{\alpha}^2 \cos^2(\delta) + \mu_{\delta}^2} =  53 \pm 11$ mas yr$^{-1}$
and PA $= 27^\circ \pm 6^\circ$ is shown by the dark cross.
The three different determinations are labeled NBPP\_2\_BON, splitted\_BON, and BON\_FITWAVE and are
described in the text. 
Solid error bars are $\pm 1\sigma$ and dashed error bars are $\pm 3\sigma$.}
\label{pm-0244}
\end{figure}

Timing noise biases proper-motion measurement unless care is taken.
Different approaches were compared, summarized in Fig. \ref{pm-0244}, where the  
labels for the methods are described below.
The first two methods compare the position at different epochs with separate data spans,
fitting for rotational parameters ($\nu, \dot\nu, \ddot\nu$) and pulsar position.
The old NBPP timing data yield a position at MJD 51250, 
with another position at MJD 53900 from the BON data,
for a proper motion of
$\mu_{\alpha} = 13.4 \pm 8 $ mas yr$^{-1}$ and $\mu_{\delta} = 58.2 \pm 2.5$ mas yr$^{-1}$
(label NBPP\_2\_BON).
In a similar way, the BON data were divided into two equal parts.
Two positions obtained at MJDs 53700 and 54500 indicate a proper motion
of $\mu_{\alpha} = 42 \pm 6$ mas yr$^{-1}$ and $\mu_{\delta} = 55 \pm 3$ mas yr$^{-1}$
(splitted\_BON).
The two last methods use the harmonic whitening procedure proposed by
\citet{Jodrell} in their Appendix A.
An iterative process using a fit of harmonically related sinusoids \citep[``FITWAVE'' in TEMPO2,][]{Edwards06}
is applied following a strict sequence of different parameter adjustments.
This procedure was applied to both the old NBPP and the new BON timing data.
The NBPP data give $\mu_{\alpha} = 24 \pm 30$ mas yr$^{-1}$ and $\mu_{\delta} = 17 \pm 12$ mas yr$^{-1}$
(NBPP\_FITWAVE).
The new BON timing data imply
a proper motion of $\mu_{\alpha} = 62 \pm 3$ mas yr$^{-1}$ and $\mu_{\delta} = 35.6 \pm 1.8$ mas yr$^{-1}$
(BON\_FITWAVE).

The error bars shown in Fig. \ref{pm-0244} for the successive measurements sometimes do not overlap
because systematic biases due to e.g., timing noise or covariances underestimated
by the fitting routines, are larger than the statistical uncertainties.
We computed a weighted mean proper motion for PSR J0248+6021 based on all the determinations except the NBPP\_FITWAVE,
which has very large uncertainty and hence constrains little.
The mean values $\mu_{\alpha}  = 48  \pm 10$ mas yr$^{-1}$ and $\mu_{\delta} = 48 \pm 4$ mas yr$^{-1}$
were obtained by weighting by the inverse of the uncertainties.
The total is
$\mu = \sqrt{\mu_{\alpha}^2 \cos^2(\delta) + \mu_{\delta}^2} =  53 \pm 11$ mas yr$^{-1}$ and 
the celestial position angle is  PA $= \tan^{-1} (\mu_{\alpha} \cos(\delta) / \mu_{\delta} ) = 27^\circ \pm 6^\circ$.

%

PSR J2240+5832's proper motion is roughly half that of the other pulsar.
The measured value became stable only after multiple iterations of the method cited above \citep{Jodrell},
apparently due to one or two marginally detectable glitches that perturb the fits.
We obtained $\mu_{\alpha}  = -6.1  \pm 0.8$ mas yr$^{-1}$ and $\mu_{\delta} = -21.0 \pm 0.4$ mas yr$^{-1}$
($\mu =\sqrt{\mu_{\alpha}^2 \cos^2(\delta) + \mu_{\delta}^2} = 21.3 \pm 0.4$ mas yr$^{-1}$ and celestial position angle PA $= 8.7^\circ \pm 1^\circ$). The uncertainties
are statistical. To be conservative, we assume systematic uncertainties of the same magnitude as those observed for PSR J0248+6021, 
and thus obtain $\mu = 21 \pm 4$ mas yr$^{-1}$ (same fractional error) and celestial position angle PA $= 9^\circ \pm 6^\circ$ (same absolute
value).

Fig. \ref{FigW5} shows both the proper motion and polarization position angles for PSR J0248+6021. They are roughly
perpendicular (difference of $94^\circ \pm 7^\circ$). \citet{2007MNRAS.381.1625J} recall that neutron star velocity vectors and spin-axis may naturally become aligned
at birth, if the supernova ``kick'' is preferentially along the spin-axis, and the
radio emission is orthogonal to the magnetic field. That paper adds a new sample of measurements for which this scenario
appears to hold about half the time, weakening the observational evidence supporting this picture. 
For PSR J2240+5832, the difference between the velocity and polarization PA's is $48^\circ \pm 6^\circ$: hence, 
in this new sample of two pulsars, again only half seem to comply with the ``rule''.

\section{Distances and surroundings}
One goal of this work is to compare model predictions of the pulsars' gamma-ray luminosity $L_\gamma$ with observations.
Following the ``Atlas'' of \citet{Watters09} and its update \citep{AtlasII}, we define
\begin{equation}
L_\gamma = 4 \pi f_\Omega G_{100} d^2
\end{equation}
for $G_{100}$, the integral gamma-ray energy flux above 100 MeV. 
The beam correction factor $f_\Omega$ is the ratio of the gamma-ray flux radiated into all space to that along the line-of-sight,
both averaged over a neutron star rotation.
Most models depend on the inclination and viewing angles $\alpha, \zeta$ as discussed below. 
The distance $d$ is clearly important. 

\subsection{Dispersion measure, kick velocities, and rotation measures}
\subsubsection{Distance to PSR J0248+6021}
As stated earlier, the DM for PSR J0248+6021 is so large for this
line-of-sight ($l=136.9^\circ$, $b=0.698^\circ$) that the NE2001 model places it outside
the Galaxy. However, Fig. \ref{FigW5} shows extensive clouds that
add to the electron column density and bias the NE2001 distance.

The pulsar lies $0.3^\circ$ west of the center of the open cluster IC 1848, which itself lies within the giant HII region W5, the ``Soul'' nebula
in the Galaxy's Perseus arm. The cluster diameter is $0.34^\circ$ \citep{2005A&A...440..403K}. 
The pulsar was born $\mu \tau_c \lesssim 1^\circ$ southwest of its current position. 
The heliocentric distance to IC 1848 estimated from both main 
sequence fitting of the star cluster and a Galactic kinematic model converges towards 2 kpc \citep{2005A&A...440..403K}.
That reference gives no distance uncertainty\footnote{See also WEBDA database, http://www.univie.ac.at/webda/}. 
Distance uncertainties for individual stars from other studies,
e.g. using photometric measurements \citep{2009MNRAS.400.1726M} do not exceed $\pm 0.4$ kpc, and we assume $\pm 0.2$ kpc.
We now argue that the pulsar is at the distance of the open cluster, i.e., on the near side of W5.

The parallel component of the magnetic field along the line-of-sight to
PSR J0248+6021 is $B_\parallel = 1.23 RM/DM = -0.5\,\mu$G. This matches nicely the
expectation from Fig. 6 of \citet{2003A&A...398..993M} and therefore the pulsar is most likely within the Perseus arm.
From \citet{MilkyWayApJ}, the far edge of that arm is $<3.6$ kpc distant.

The proper motion of PSR J0248+6021 is, with its uncertainty, $>42$ mas yr$^{-1}$.
The largest measured neutron-star velocities due to supernova ``kicks'' at birth are
of the order of 1000 km s$^{-1}$ \citep{HobbsProperMotions}. Thus, if PSR J0248+6021 were beyond $4.8$ kpc it would have an unusually
large kick velocity. 
Conversely, a 2 kpc distance implies a very typical transverse 
velocity of $v_T \sim 500$ km s$^{-1}$.
Furthermore, the DM of PSR J0248+6021 increased steadily by $0.3$ cm$^{-3}$ pc per year from 2006 to 2008,
and decreased at the same rate in 2009 and 2010. Fig. \ref{ZoomedResiduals} shows that the detailed
behavior is complex.
The simplest interpretation of the DM evolution is
that the pulsar is moving rapidly in a dense environment, that is, that the pulsar is in a cloud
(conceivably a nearer, invisible cloud).

For the cloud coincident with the pulsar (Fig. \ref{FigW5}),
the largest velocity (largest Doppler shift) of reliable HI features towards PSR J0248+6021 is $-50$ km s$^{-1}$.
In light of the known velocity reversal in this part of the Perseus arm \citep{2010MNRAS.tmpL..26T} the cloud
is thus between 2 and 3 kpc, like the rest of W5.
High resolution H$\alpha$ imaging could confirm this preliminary indication. 

Towards PSR J0248+6021, the NE2001 model places the edge of the stellar Galactic disk 9 kpc from
the Sun, with a DM of $210$ cm$^{-3}$ pc.
A dense local environment, as within W5, may explain
the large observed electron column density DM $= 370$ cm$^{-3}$ pc.
For 2 kpc, NE2001 predicts DM $\simeq 61$ cm$^{-3}$ pc, implying an excess of $\sim$ 315 cm$^{-3}$ pc. 
The main part of the HII region neighboring the pulsar is a shell of diameter $0.5^\circ$ on the sky, 
equivalent to a true size of 17.5 pc at 2 kpc.
This converts into an electron density in the cloud of $<n_e> \simeq 18$ cm$^{-3}$. 
Though large, this is typical of other HII regions, where 
values from 0.1 to 100 cm $^{-3}$ are seen (see e.g. \cite{2001A&A...370..586M}, 
for their study of the scatter broadening in the direction of the Gum Nebula).
The WISE collaboration \citep{2010AAS...21610401W} released a detailed image
of the Heart and Soul nebulae seen between $3.4$ and 22 microns\footnote{http://antwrp.gsfc.nasa.gov/apod/ap100601.html}.
PSR J0248+6021 lies in one of the ``thickest'' areas, where the shell is seen edge-on and
the electron density is surely greater than the average value, indicating again that the large
DM is compatible with a small pulsar distance.

The large scattering tail excludes the notion that PSR J0248+6021 may be within the Orion spur, thus
only $\sim 1$ kpc away, because of nearby but invisible electron clouds.
To conclude, PSR J0248+6021 is almost certainly in W5, at the distance of IC 1848, and we use $d = 2.0 \pm 0.2$ kpc.

\subsubsection{Distance to PSR J2240+5832}
PSR J2240+5832 has DM $= 263.5$ cm$^{-3}$ pc and along this line-of-sight ($l=106.57^\circ$, $b=-0.111^\circ$) the
NE2001 distance is $10.3^{+\infty}_{-3.3}$ kpc.  
Archival optical and radio images show no bright structures near the pulsar position that would indicate
significant deviations from the NE2001 maps.
However, the pitch angle of the outer arm
has been shown to be smaller than known when the NE2001 model was created \citep{MilkyWayApJ}, 
and the far edge of the outer arm is $< 8.3$ kpc in this direction. 

Again assuming a transverse velocity $v_T<1000$ km s$^{-1}$, the proper motion of $>17$ mas yr$^{-1}$ 
of PSR J2240+5832 limits its distance to $<12$ kpc.
Conversely, a value of $v_T = 400$ km s$^{-1}$ would place the pulsar in the Perseus arm, which
extends from 3 to $4.6$ kpc in this direction. 
PSR J2240+5832 is therefore among the few pulsars in this direction for which
rotation measures are available \citep{HanRotationMeasures}. 
The positive value of RM$=24$ rad m$^{-2}$, albeit small, is anomalous:
other low-latitude RM's for $90^\circ < l < 120^\circ$ are negative. 
On the other hand, in this direction no other pulsar between 6 and 8 kpc has an RM measurement
and the positive value may be due to the
magnetic field inversion beyond the Perseus arm suggested by \citet{HanRotationMeasures}.
The magnetic field $B_\parallel = 1.23 RM/DM = 0.1\,\mu$G is within the range of observed values
in \citet{2003A&A...398..993M}.

Neutron stars are most likely born within arms and, since $v_T \tau_c < 0.1$ kpc,
the pulsar should not have gone far from its birthplace.
We conclude by presenting two possible values for the distance to PSR J2240+5832: 
one in the outer arm ($d=7.7 \pm 0.7$ kpc) and another in the Perseus arm ($d=3.8 \pm 0.8$ kpc).
If in the outer arm, $v_T$ is large but not exceptional and the pulsar is
farther than most known gamma-ray pulsars.
If it is in the Perseus arm, the DM excess and positive RM remain unexplained but the
speed and distance are more typical. We see below that the two values of $L_\gamma$ 
obtained using the two distances are both plausible.
VLBI astrometric parallax measurements would be of great interest for these and all gamma-ray pulsars.

\subsection{Evidence of a pulsar wind nebula?}
PSR J0248+6021 coincides with the heart of an elongated nodule $15^\prime$ long (see Fig. \ref{FigW5}), 
seen at both 21 cm \citep[NVSS,][]{1998AJ....115.1693C} and in the optical. 
The pulsar proper motion is $\sim 45^\circ$ from the nodule long axis. 
The POSS-II (Palomar Optical Sky Survey) image shows an arc $20^{\prime\prime}$ long about $5^{\prime\prime}$ from the pulsar position.
The simple formulae for the size and offsets for pulsar wind nebula (PWN) termination shocks given
by \citet{2008AIPC..983..171K} are roughly consistent with what we observe. 
Might we be seeing shocks in the interstellar medium caused by the pulsar's high transverse velocity?

To explore the idea of a possible PWN,
we obtained \textit{Swift} \citep{2004ApJ...611.1005G} observations of PSR~J0248+6021 with data taken 
over two days, 2009 Dec. 10-11. The target was not detected in the X-ray 
Telescope \citep[XRT;][]{2005SSRv..120..165B} combined 7.2 ks exposure and we
derive a $3 \sigma$ limit on the source count rate (0.5--8 keV) of $0.0024$ cts/s 
after dead time, vignetting, and PSF corrections. Using PIMMS, this corresponds to 
an observed flux limit of $1.4 \times 10^{-13}$ erg cm$^{-2}$ s$^{-1}$, 
or $2.0\times 10^{-13}$ erg cm$^{-2}$ s$^{-1}$ unabsorbed with Galactic 
$N_H = 8.3\times 10^{21}$ cm$^{-2}$ \citep{2005A&A...440..775K}, 
assuming a power-law photon flux with $\Gamma=1.5$ as is typical
among observed X-ray PWN (same sign convention for $\Gamma$ as in Eq. 2, Sect. 5).
For $d=2$ kpc, we obtain (absorbed) $L_X < 6\times 10^{31}$ erg/s for the PWN X-ray luminosity.
Only eight of the 40 X-ray PWN's in \citet{2008AIPC..983..171K} are dimmer: if there is an X-ray PWN, it is faint,
possible for a PWN of the age of these pulsars.

An upper limit on GeV PWN emission is given in Section 5.1.
The position of PSR J0248+6021 is at the corner of both the MAGIC \citep{MagicLSI} 
and the VERITAS \citep{VeritasLSI} TeV significance maps for LS I$+61^\circ$ 303.
No evidence of TeV emission from a putative PWN is apparent.
Reanalysis of the complete TeV datasets optimized for a source at this location 
with PWN-like properties (i.e. possibly offset, and/or spatially extended) is encouraged.

%

\subsection{Upper limit for detection of an optical pulsar}
Digital Sky Survey images show a faint source seemingly coincident with the pulsar position.
Analyzing the Ultra-Violet/Optical Telescope \citep[UVOT;][]{2005SSRv..120...95R} U-filter image (exposure: 3575 s) obtained 
with \textit{Swift} on 2009 December 11th, the faint source is detected at $6.0\sigma$.
We measure its center (J2000) at RA = 02h 48' 18.711", 
dec = 60d 21' 38.98" with a statistical uncertainty of 0.46" (90\% confidence). 
Its observed flux density is $1.07 \pm 0.18$ $\times 10^{-2}$ mJy 
at $8.56 \times 10^{14}$ Hz corresponding to magnitude $20.33 \pm 0.18$.
The uncertainties are statistical. 
This source is not detected in the UVOT W1-filter image (3537 s) obtained on 2009 December 10th and we derive a $3\sigma$ upper limit
of $< 2.80 \times 10^{-3}$ mJy at $1.14 \times 10^{15}$ Hz. 
Pulsar timing is incompatible
with the position of the optical object and there is therefore no evidence of an optical pulsar.

\section{Gamma-ray Observations}\label{sec:gamobs}
The LAT is an electron-positron pair conversion telescope that was placed in orbit with the \textit{Fermi} satellite 
on 2008 June 11 \citep{LATinstrument,OnOrbit}. 
The LAT covers the 20 MeV to $>300$ GeV energy range with higher sensitivity and more accurate localisation than previous
instruments (an on-axis effective area $\sim 8000$ cm$^2$ above 1\,GeV and angular resolution $\theta_{68} \sim 0.8^\circ$ at
1\,GeV).
LAT measurements yielded a catalog of 1451 gamma-ray ``1FGL'' sources during its first year \citep{1FGL}. 

We used the standard \textit{Science Tools} software package for \textit{Fermi} LAT data analysis\footnote{Gamma-ray data, 
analysis software, rotation ephemerides, and the diffuse background models are publically available 
at the {\em Fermi} Science Support Center, FSSC, http://fermi.gsfc.nasa.gov/ssc/.}.
For both pulsars, we selected data collected between 2008 August 4 (MJD 54682) when \textit{Fermi} began scanning-mode operations,
and the end of the ephemeris validity range (over 15 months in both cases). 
We kept ``diffuse'' class events (highest probability of being $\gamma$-ray photons) within a $15^\circ$ 
``region-of-interest'' (ROI) around the pulsar. 
We excluded events with zenith angles $>105^\circ$ to reject the $\gamma$-ray albedo from the Earth's limb.
Photon phases were calculated using the TEMPO2 pulsar timing software \citep{Edwards06}. 

At low photon energies, multiple scattering dominates the LAT's angular resolution.
For the gamma-ray pulse profiles (phase histograms), we applied an energy-dependent angular radius cut centered on
the pulsar that approximates the instrument \emph{point spread function} (PSF), given by 
$<\theta_{68}(E)> = \sqrt{ (5.12^\circ)^{2}(100 \ {\rm MeV} /E)^{1.6} + (0.07^\circ)^{2} }$,  
which corresponds to a 68\% containment angle (\textit{P6\_V3} Instrument Response Function, ``IRF'').
A maximum radius of $\theta_{68}^{\rm max} = 0.8^\circ$ and $0.7^\circ$ for PSRs J0248+6021 and J2240+5832, 
respectively, reduces the background at low energies. We also remove events
from nearby 1FGL sources within $<3^\circ$ using the same bounded, energy-dependent radius as applied to the pulsar.

The on-pulse spectra were obtained with a maximum likelihood analysis \citep{1996ApJ...461..396M} of the LAT data within $15^\circ$
of the pulsar, using the \textit{Fermi} science tool ``gtlike''. The likelihood method weights events from the target and 
background sources according to the energy-dependent PSF, which explains the need for such a large ROI. 
We excluded time intervals when the ROI intersected the Earth's limb. 
In this paper, we modeled the pulsars using the functional form
\begin{equation}\label{eq:spectra}
{dN \over dE} = N_0 E^{-\Gamma} {\rm exp} \left[-\left(\frac{E}{E_{c}}\right)^{\beta}\right]\, {\rm cm}^{-2}\, {\rm s}^{-1}\, {\rm MeV}^{-1}, 	
\end{equation}
where $N_0$ is the differential flux normalization (ph\,cm$^{-2}$\,s$^{-1}$\,MeV$^{-1}$), $\Gamma$ the photon index, 
and $E_{c}$ the cutoff energy. A pure power law ($\beta = 0$) describes many gamma-ray sources, such as
active galactic nuclei \citep{1FGL}. 
The 46 gamma-ray pulsars discussed in \citet{FermiPsrCata} are generally well-described by a simple exponential cutoff,
$\beta=1$, a shape predicted by outer magnetosphere emission models (see the Discussion, below). 
Models where gamma-ray emission occurs closer to the neutron star can have sharper ``super-exponential'' cutoffs,
e.g. $\beta = 2$. 

The Galactic diffuse emission was modeled using the
\textit{gll\_iem\_v02} map cube based on six Galactocentric ``ring'' maps of N(H$_I$) and W(CO) and on the spatial 
distribution of the inverse Compton intensity modeled by GALPROP \citep{Strong2004a,Strong2004b}.  
Diffuse extragalactic gamma-ray emission and residual instrument backgrounds were modeled jointly by the isotropic 
component \textit{isotropic\_iem\_v02}. Both models are available from the FSSC. All 1FGL sources within $20^\circ$ 
were modeled with a power-law. Sources farther than 5$^\circ$ from the target pulsar were assigned fixed power law spectra, 
with parameters taken from the 1FGL source catalog. Spectral parameters for the pulsars and sources within 
5$^\circ$ of the pulsar were allowed to vary. 
Systematic uncertainties were estimated by reapplying the fitting procedures using
bracketing IRFs where the effective area was shifted (linear extrapolations in log space) 
by $\pm$ 10\% at 0.1 GeV, $\pm$ 5\% near 0.5 GeV, and $\pm$ 20\% at 10 GeV.

\subsection{PSR~J0248+6021}
PSR~J0248+6021 is called 1FGL~J0248.3+6021 in the LAT catalog \citep{1FGL}.
It is in the Galactic plane, $1.3^\circ$ away from the bright gamma-ray source 1FGL~J0240.5+6113, 
identified as the Be star binary LSI +61$^\circ$ 303 \citep{LATLSI+61}. No other 1FGL sources
are within $3^\circ$.  
The high $\dot E$ pulsar PSR~J0205+6449 in 3C~58 is about $6^\circ$ away \citep{LATPSR0205}. 
PSR~J0248+6021 was discovered towards the end of EGRET's
lifetime, and ephemerides contemporaneous with only a fraction of EGRET data exist. 
Before the \textit{Fermi} mission, we searched
for gamma-ray pulsations in all EGRET data for a range of ($\nu$, $\dot \nu$) values extrapolated from our radio ephemeris. However,
no pulsed signal was detected presumably because of the poor S/N of the EGRET data.

\emph{Fermi} LAT data for PSR~J0248+6021 do however exhibit clear pulsations  
for the 3810 photons with energies $> 0.1$ GeV remaining after cuts.
Fig. \ref{phasos_0248} shows the folded light curve of these events for different energy bands, along
with the radio profiles used to derive the timing parameters. 
The light curve in each energy interval consists of
one broad peak between 0.20 and 0.55 in phase, which dominates the highest energy band ($> 1$\,GeV) but is
progressively less pronounced with decreasing energy. 
The pulsation significances based on the bin-independent \textit{H}-test \citep{DeJager2010} 
are $\sim 1\sigma$, $6.5\sigma$, $8.2\sigma$, respectively, for the three energy bands $0.1-0.3$, $0.3-1.0$, and $>$1\,GeV.
Between $0.3$ and 1\,GeV, a structure with statistical significance of $\sim$3
$\sigma$ (signal/$\sqrt{\rm background}$) appears between phases 0.6 and 0.7. 
However, it is not observed in the other energy bands, and a full likelihood analysis
in that phase interval yields an excess with a significance of only $2\,\sigma$
between 0.1 and 100\,GeV. 
We fit the main $\gamma$-ray peak ($0.2 < \phi < 0.55$) above 0.1 GeV with two
half-Lorentzian functions. This shape provides a higher value of $\chi^2$ than others and accommodates differences
between the leading and trailing edges. The fit peak is at phase $\delta = 0.39 \pm 0.02 $
after the maximum of the radio peak, with a full width at half maximum (FWHM) of $0.20 \pm 0.02$. 
The uncertainty in $\delta$ is statistical. The bias due to the DM uncertainty in extrapolating the radio 
TOA to infinite frequency is negligible.
The peak position does not vary with $\gamma$-ray energy within the statistical uncertainties, and the highest
energy photon has 5\,GeV at phase $0.31$.

We fit the on-pulse ($0.2<\phi<0.55$) data for PSR~J0248+6021 with an exponentially cutoff power law ($\beta = 1$).
The two neighbors mentioned above, as well as PSR~J0218+4232 \citep{MSP}, were modeled in similar ways.
Softer or sharper cutoffs ($\beta \neq 1$) yield essentially the same statistical significance.
The likelihood ratio test ($2\Delta(loglike)$, twice the difference of the logarithm of the likelihood maximum value) 
prefers the simple exponentially cutoff power law ($\beta = 1$) for PSR~J0248+6021 to a simple
power law ($\beta = 0$) by $8 \sigma$. The spectral results for $\beta =1$ are listed in Table \ref{Table}, 
along with the integral photon flux $F_{100}$ and integral energy flux $G_{100}$ above $0.1$ GeV.
The first errors are statistical and the second are systematic. 
Fig. \ref{J0248gammaSED} shows both the overall fit between 0.1 and 9\,GeV
(solid lines) with $\beta = 1$, and spectral points from likelihood fits in each individual energy band
using a power-law spectrum. 

We performed a maximum likelihood analysis for 50 days before and after the 2009 March glitch epoch. 
No change in the flux was observed.
We searched the off-pulse data for a point source in the energy band 0.1-100 GeV at the radio pulsar
position. No signal was observed. After scaling to the full pulse phase, we derived a
95\% confidence level upper limit on the flux of $1 \times 10^{-8}$\,cm$^{-2}$ s$^{-1}$. 
If a PWN is associated with the pulsar, deep observations will be required to see it.

\subsection{PSR J2240+5832}
PSR~J2240+5832 is in the Galactic plane $0.6^\circ$ away from the radio-quiet gamma-ray pulsar 
PSR~J2238+5903  \citep[$S_{1400} < 7\,\mu$Jy,][]{BlindTiming}, and $3^\circ$ away from the Vela-like PSR~J2229+6114 
associated with the ``Boomerang'' PWN and SNR \citep{LATVelaLike}. 
No other 1FGL source lies within $3^\circ$. 
The high background from nearby PSR~J2238+5903 is the reason for the choice of small value of $\theta_{68}^{\rm max} = 0.7^\circ$.
A total of 1208 photons with energies $>0.3$\,GeV were selected. 
Fig. \ref{phasos_2240} shows the gamma-ray pulse profiles in different energy bands, 
phase-aligned with the peak of the $1.4$ GHz radio profile observed at Nan\c{c}ay. 
A single peak between phases $0.55$ and $0.63$ appears above $0.3$ GeV only. 
We tried different apertures to search for a signal below $0.3$\,GeV, but no excess was observed at the peak location. 
The highest \textit{H}-test significance is 6.3\,$\sigma$ and occurs above 1\,GeV.
Fitting above $0.3$ GeV with a Lorentzian function places the peak at $\delta = 0.58 \pm 0.01 $
after the radio maximum, with an FWHM of $0.11 \pm 0.02$. The peak position does not vary with $\gamma$-ray energy 
within statistical uncertainties, and the highest energy photon (20\,GeV) lies at $\phi = 0.72$, 
just outside the peak phase range.

\begin{figure}[htbp]
\begin{center}
\epsfig{file=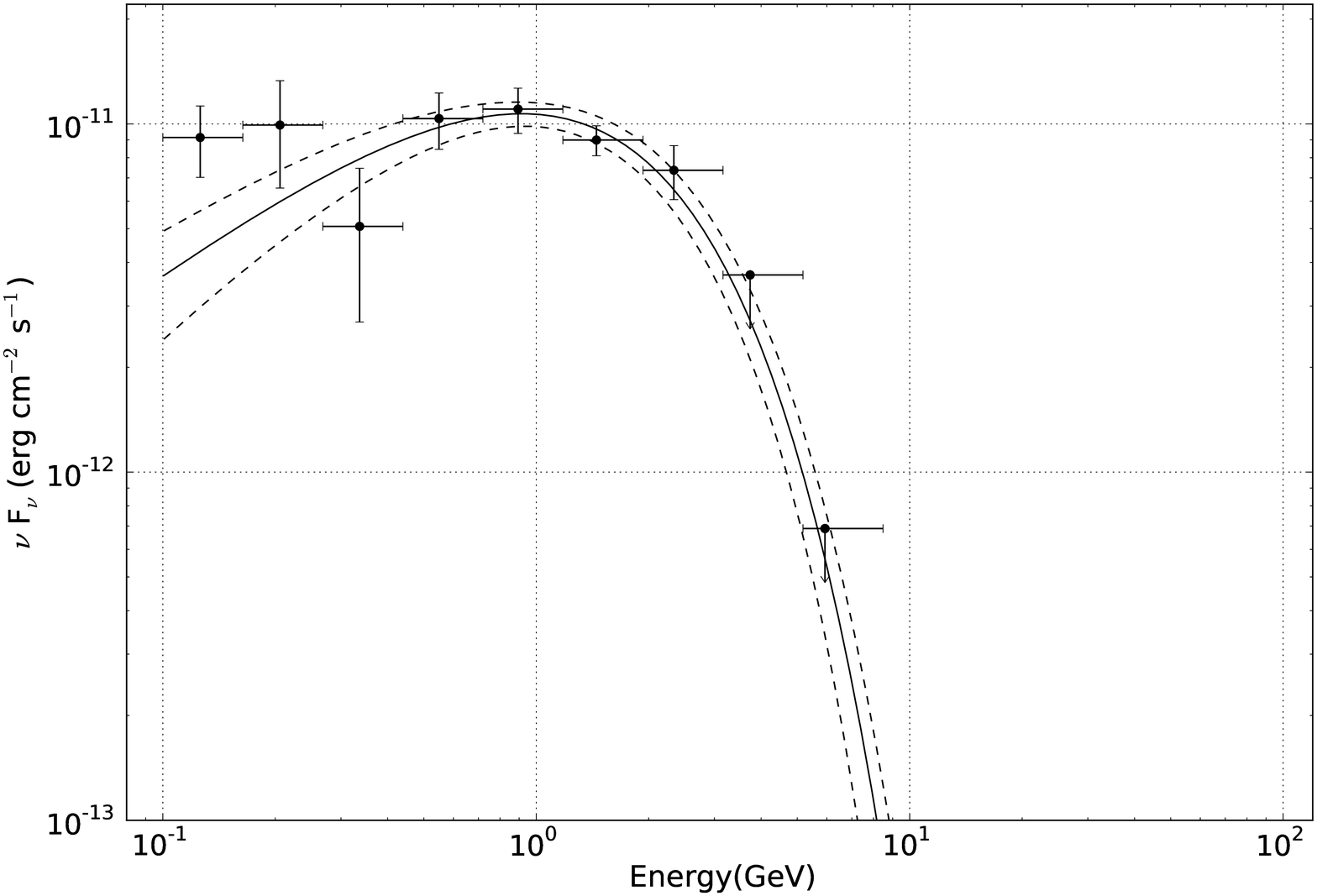,width=9.0cm}
\epsfig{file=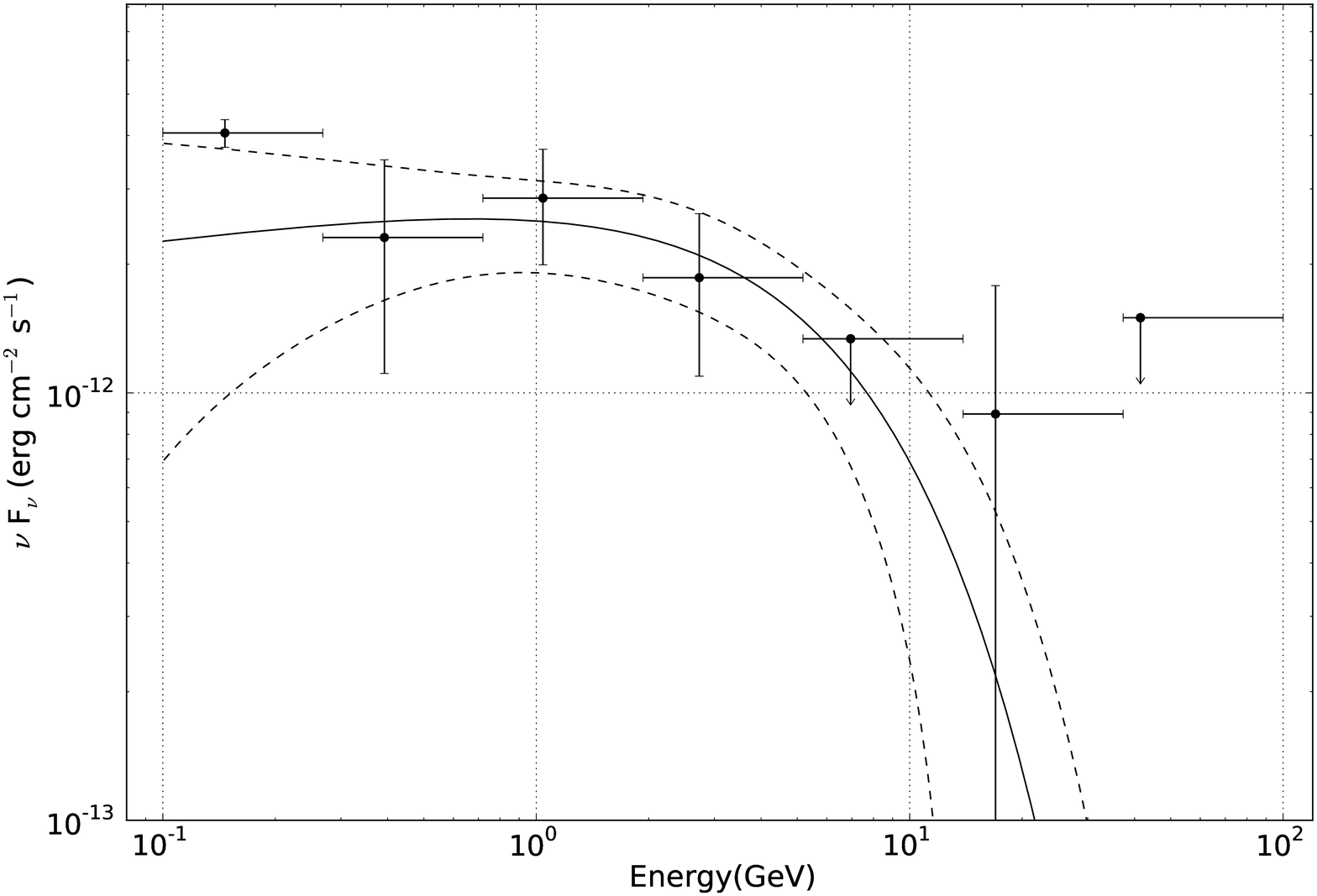,width=9.0cm}
\caption{On-pulse gamma-ray spectral energy distributions (SEDs) obtained with the \emph{Fermi} Large Area Telescope. 
{\em Top}: PSR J0248+6021, in the phase range $0.2<\phi<0.55$.
{\em Bottom}: PSR~J2240+5832 in the phase range $0.50<\phi<0.75$.
Plotted points are from power-law likelihood fits to individual energy bands with $\geq 2 \sigma$ detection significance 
above background for two degrees of freedom, otherwise an upper limit arrow is shown.
The solid black line shows the maximum likelihood fit to a power law with
exponential cutoff (Eq. 2). The dashed lines are $\pm 1 \sigma$ uncertainties in the fit parameters. 
}
\label{J0248gammaSED}
\end{center}
\end{figure}

We determined the on-pulse ($\phi = 0.50 - 0.75$) spectrum using the
maximum likelihood analysis within a 15$^\circ$ aperture. Both nearby pulsars PSRs~J2238+5903 and J2229+6114 
as well as PSR~J2240+5832 were modeled by a simple exponentially cutoff power law ($\beta = 1$ in Eq. 2). 
Results are listed in Table \ref{Table}. 
The source is three times fainter than the other pulsar, leading to large uncertainties 
in the spectral parameters. The spectral index and cutoff are consistent with those seen for most young gamma-ray pulsars.
We explored different models and the exponential cutoff ($\beta=1$) 
is preferred to a power law by only $2 \sigma$.  Fig. \ref{J0248gammaSED} shows both the overall
fit between 0.1 and 100\,GeV (solid lines), along with the spectral points from power-law likelihood fits to each
individual energy band. 

As for PSR J0248+6021, we searched the off-pulse region for a possible PWN at the pulsar position. 
No signal was observed.
Scaling to the entire phase range (a full rotation) yields
a 95\% confidence level upper limit on the flux of $2\times 10^{-7}$ cm$^{-2}$ s$^{-2}$. 
We note however that
the MILAGRO significance map for the region surrounding the ``Boomerang'' PWN and PSR J2229+6114 extends intriguingly
towards the Galactic plane, reaching $4.7\sigma$ at the position of PSR J2238+5903 \citep{MilagroFermiBSL}.
The Tibet air shower array similarly sees a $2.5\sigma$ excess at that location \citep{Tibet}. Both
MILAGRO's and Tibet's angular resolutions are poor, and TeV PWN's are notoriously offset from the
pulsars driving them. PSR J2238+5903 is more energetic ($\dot E = 9 \times 10^{35}$ erg s$^{-1}$) than PSR J2240+5832
and presumably closer, given that it is brighter ($F_{100} = 6.8 \pm 1.5 \times 10^{-8}$ photons cm$^{-2}$ s$^{-1}$ above 100 MeV). 
The TeV excess might be a conflation of two adjacent sources. Deeper observations with a Cherenkov imager 
array may be interesting.

\section{Discussion}
\subsection{Gamma-ray luminosity}
Armed with the gamma-ray integral energy flux $G_{100}$ and the distances $d$, 
we can now evaluate the luminosity $L_\gamma$ (Eq. 1).
We set $f_\Omega=1$ as in \citet{FermiPsrCata}, and see below that the
emission models in any case yield $f_\Omega \simeq 1$.
Defining $G_{11} = 10^{-11}$ erg cm$^{-2}$ s$^{-1}$ and $d_1 = 1$ kpc $=3.1 \times 10^{21}$ cm gives
$L_\gamma = 1.2 \times 10^{33} f_\Omega (G_{100}/G_{11}) (d/d_1)^2$ erg s$^{-1}$.
For PSR J0248+6021, summing the distance and $G_{100}$ uncertainties in quadrature
yields $L_\gamma = (1.4 \pm 0.3) \times 10^{34} f_\Omega $ erg s$^{-1}$.
For PSR J2240+5832, the distance ambiguity leads to larger uncertainties: if the pulsar
is in the outer arm, we obtain $L_\gamma = (7.9\pm 5.2) \times 10^{34} f_\Omega $ erg s$^{-1}$, but if there
is some unseen cloud along the line-of-sight creating an overdensity of electrons,
and the pulsar is in the nearer Perseus arm, then $L_\gamma = (2.2 \pm 1.7) \times 10^{34} f_\Omega $ erg s$^{-1}$.
The corresponding efficiencies $\eta = L_\gamma/\dot E$ are listed in Table 1.

We compare these luminosities with the rule-of-thumb for young pulsars illustrated in Fig. 6 of 
the {\em Fermi} Pulsar Catalog \citep{FermiPsrCata}, 
$L^h_\gamma = 10^{33}\sqrt{{\dot E}/10^{33}}$ erg/s, where $h$ stands for ``heuristic''. 
For the $\dot E$ of our two pulsars, $L^h_\gamma$ is $1.5 \times 10^{34}$ erg s$^{-1}$, 
very near $L_\gamma$ obtained for PSR J0248+6021. 
The measured value of $L_\gamma$ for PSR J2240+5832 overlaps the rule-of-thumb for the distance of
the Perseus arm, while the outer arm value is within the spread of values observed for the other pulsars.
It is important not to assign a distance to the pulsar based on a comparison of observed $L_\gamma$
with expectations, since that may bias future attempts to use population modeling to refine
emission models.

\subsection{Emission models}
These two pulsars have atypical gamma-ray pulse profiles: 
as for the six pulsars studied by \citet{LAT6pulsars}, they have only one gamma-ray peak, and
like four of those, they have a large offset from the radio pulse. 
Most gamma-ray pulsars have two peaks, the first lagging the radio beam by $<0.2$ rotations. 
Hence, these two pulsars help extend the parameter space over which we can test 
gamma-ray emission models. 

To pin down the geometrical angles $\alpha$ and $\zeta$, beyond the RVM work described above,
we follow \citet{AtlasII} in modeling the gamma-ray pulse profiles. Byproducts
are the flux correction factors $f_\Omega$, which reduce the uncertainty in $L_\gamma$.
We test two simple versions of outer magnetosphere pulse models. The first
model \citep[`outer gap', or OG, ][]{OuterGap} has emission starting at the `null charge' surface and extending
to the light cylinder. The second model \citep[`two pole caustic', or TPC,][]{TwoPoleCaustic} 
has emission starting at the star surface and extending to a perpendicular
distance of $0.75R_{LC}$ from the rotation axis or a radial distance of $R_{LC}$, whichever is less. 
In both cases, the emission comes from a zone spanning a characteristic 
fraction $w = L^h_\gamma/\dot E = 0.075$ of the open zone near the last-closed field lines. 
For PSR J0248+6021, we choose a linear intensity gradient across the gap but the modeled light
curve shape is not very sensitive to the illumination across the gap zone,
and narrow or uniformly illuminated gaps give similar results. 
In contrast, the very sharp $\gamma$-ray pulse of PSR J2240+5832
seems to require a relatively small range of field lines near the
maximum $w$ to dominate the emission: here we used a simple Gaussian
weighting, peaked around $w=0.075$.
We also plot the fits to the numerical plasma-filled (force free) model
of \citet{SeparatrixLayer}.
This so-called `separatrix layer' (SL) model
has a dense plasma and currents and posits emission from field lines
extending well outside the light cylinder into the wind zone.

Figs. \ref{J0248_OGTPC} and \ref{J2240_OGTPC} show the goodness-of-fit surfaces in the
$(\alpha, \zeta)$ plane for the two pulsars.
``Goodness-of-fit'' matching of the observed and modeled post-profiles uses
the exponentially-tapered $\chi_3$ weighting defined in \citet{AtlasII}, a more
robust test statistic than a simple $\chi^2$. 
The regions selected by the RVM fits to the radio data are superimposed. 
In both cases, emission from near the light cylinder (as used in this OG model) seems necessary to provide good fits. 
In addition, emission from below the `null charge' surface appears to
create too many pulse components and too much off-pulse emission.
For PSR J0248+6021, the best-fit curves in the radio-allowed region
are near $\alpha=46^\circ$, $\zeta=52^\circ$. This model produces
a broad, merged peak at the correct phase with a good 
fit statistic $\chi_3=2.4$. The inferred flux correction factor is
$f_\Omega =1.06$. The best fit TPC model in the allowed region
is near the same angles but gives a substantially poorer fit, 
because of a second pulse component at $\phi \sim 0.1$. Here we
have $f_\Omega= 0.86$.

A similar exercise for PSR J2240+5832 shows rather similar regions of best fit for the two models. 
However, again the OG model provides a better fit (near $\alpha = 101^\circ$,
$\zeta=117^\circ$) with $\chi_3=4.1$. The poorer TPC fit at this position has $\chi_3=7.5$. 
In this case, the OG light curve is dominated by a single narrow pulse near $\phi=0.6$.
A much weaker component near $\phi=0.2$ is faint or missing in the present gamma-ray light curve. 
Although these same two components are present in the TPC curve, the first peak after the radio
pulse is comparably bright to the second, and for these $\alpha, \zeta$ angles significant emission 
is predicted for all pulsar phases. 
These differences indicate that, using the particular realizations
of the OG and TPC models described by \citet{AtlasII}, the data are most consistent with the OG-type picture. 
For both fits, $f_\Omega \approx 1$. 
The SL models provide $\chi_3$ fit levels intermediate between
the OG and TPC models. The relatively good fits tend to lie along
a narrow band and the intersection with the RVM-allowed angles is similar
to that proposed in the OG case.

Thus, while both of these pulsars show single dominant components in the GeV light curves, the pulse profile modeling
suggests a different origin. For PSR J0248+6021, the single broad component
near phase $\phi\approx 0.3-0.45$ may be a merged double from
a grazing sweep across the hollow $\gamma$-ray beam. In contrast,
the very narrow pulse at phase $\phi=0.6$ for PSR J2240+5832
implies a cut through a caustic surface for a component
normally identified with the second peak in young pulsars: 
the first peak is weak or absent.

\section{Conclusions}
The Nan\c cay survey of the northern Galactic plane led to the discovery of two remarkably similar young pulsars. 
They were subsequently timed for twelve years, during which period instrument precision
improved steadily. The extensive timing data
allowed proper motion determinations, and revealed a slowly varying DM for one pulsar. These data, together with RM measurements
obtained with polarization data, proved crucial in constraining the distance of PSR J0248+6021 to 2 kpc, much
less than the value deduced from the DM alone. The timing data also facilitated investigations of the pulsar's environment.
Radio polarization data constrained the pulsar geometry.

Both pulsars have large spin-down power and are among the growing number of gamma-ray pulsars detected 
with the {\em Fermi} Large Area Telescope. Yet they are in the minority of gamma-ray pulsars
displaying only a single peak, with a large offset from the radio pulse. Gamma-ray emission modeling
constrains the regions where the electrons radiate at high energies.
The gamma-ray luminosities obtained reinforce the growing evidence for how emission depends
on spin-down power.

We have studied radio, optical, X-ray, and TeV data for evidence of either accompanying pulsar wind nebulae or a
bright optical pulsar. None was seen, but we presented evidence to encourage deeper X-ray and TeV searches.
Sustained, accurate pulsar timing combined with multi-wavelength data are thus allowing progress in the understanding
of the emission mechanisms at work around neutron stars, and of the diffuse Galactic radiations at all
energies.

\begin{table*}
\caption{Measured and derived parameters for pulsars PSR~J0248+6021 and PSR~J2240+5832.}
\begin{tabular}{lll}
\hline\hline
Pulsar name\dotfill & J0248+6021 & J2240+5832\\ 
Right ascension,  (J2000)\dotfill &  02:48:18.617(1) & 22:40:42.939(4) \\ 
Declination, (J2000)\dotfill & +60:21:34.72(1) & +58:32:37.52(4)\\ 
Position epoch, (MJD)\dotfill & 54000 & 54000 \\
Galactic longitude, $l$ (degrees)\dotfill & 136.9 & 106.6\\ 
Galactic latitude, $b$(degrees)\dotfill & 0.698 & -0.111\\ 
Pulse frequency, $\nu$ (s$^{-1}$)\dotfill & 4.6063 & 7.1462\\ 
First derivative of pulse frequency, $\dot{\nu}$ ($10^{-12}$ s$^{-2}$)\dotfill & $-1.169$& $-0.7843$ \\ 
Proper motion in right ascension, $\mu_{\alpha}$ (mas\,yr$^{-1}$)\dotfill & $48 \pm 10$ & $-6.1 \pm 0.8$\\ 
Proper motion in declination, $\mu_{\delta}$ (mas\,yr$^{-1}$)\dotfill & $48 \pm 4$ & $-21.0 \pm 0.4$\\ 
Timing data span (MJD)\dotfill & 54682.7 to 55160.9 & 54682.7 to 55202.6\\ 
Rms timing residual ($\mu s$)\dotfill & 181 & 66\\
Radio pulse width at 1.4 GHz, $W_{50}$ (phase) \dotfill & 0.12 & 0.07 \\
Dispersion measure, $DM$ (cm$^{-3}$pc)\dotfill & $370 \pm 1$ & $263.50 \pm 0.05$ \\ 
DM epoch, (MJD)\dotfill & 54000 & 54000 \\
Rotation measure, $RM$ (rad m$^{-2}$)\dotfill & $-158 \pm 6$ & $24 \pm 4$ \\ 
Mean flux density at 1400 MHz, $S_{1400}$ (mJy) \dotfill & $13.7 \pm 2.7$  & $2.7 \pm 0.7$\\
Radio spectral index \dotfill & $-1.2 \pm 0.1$ & $-1.8 \pm 0.3$ \\
Radio-gamma-ray peak offset, $\delta$ \dotfill & $0.39 \pm 0.02$ & $0.58 \pm 0.01$\\
\hline
\multicolumn{3}{c}{Gamma-ray spectral Parameters} \\
Differential photon flux normalization, $N_0$ (cm$^{-2}$ s$^{-1}$ MeV$^{-1}$)  \dotfill & $1.5 \pm 0.7 \pm 0.1$ & $0.13 \pm 0.06 \pm 0.10$ \\
Integral photon flux, $F_{100}$ ($> 100$ MeV, $10^{-8}$ ph cm$^{-2}$ s$^{-1}$) \dotfill & $4.0 \pm 0.4 \pm 0.5$ & $1.5 \pm 0.8 \pm 0.4$\\
Integral energy flux, $G_{100}$ ($> 100$ MeV, $10^{-11}$ erg cm$^{-2}$ s$^{-1}$) \dotfill & $2.9 \pm 0.3 \pm 0.2$ &$1.0 \pm 0.4 \pm 0.2$ \\
Exponential cut-off energy, $E_c$ (GeV) \dotfill  & $1.2 \pm 0.2 \pm 0.1$  & $5.7\pm 4.4 \pm 1.0$\\
Power law index, $\Gamma$  \dotfill  &$ 	1.3 \pm 0.1 \pm 0.1$  &$1.8 \pm 0.6\pm 0.1$ \\
\hline
\multicolumn{3}{c}{Derived Quantities} \\
Spin-down power, $\dot E$ (erg s$^{-1}$) \dotfill & $2.13 \times 10^{35}$ & $2.12 \times 10^{35}$ \\
Characteristic age, $\tau_c$ (kyr) \dotfill & 63 & 151 \\
Surface magnetic field strength, $B_S$ ($10^{12}$ G) \dotfill & 3.5 & 1.45\\
Magnetic field strength at the light cylinder, $B_{LC}$ (G) \dotfill & 3150 & 4860 \\
Distance from NE2001, $d_{NE}$ (kpc) \dotfill & $>43.5$ & $10.3_{-3.3}^{+\infty}$ \\
Distance used in this work, $d$ (kpc) \dotfill & $2.0 \pm 0.2$ & $7.7 \pm 0.7$ or $3.8 \pm 0.8$ \\
Luminosity, $L_\gamma$ ($E > 100$ MeV, $10^{34}$ erg s$^{-1}$) \dotfill & $1.4 \pm 0.3$ & $7.9\pm 5.2$ or $2.2 \pm 1.7$\\
Efficiency, $\eta = L_\gamma/\dot E$  (\%)\dotfill  & $7 \pm 2$ & $37 \pm 25$ or $10 \pm 8$\\
\hline
\end{tabular}
\label{Table}
\end{table*}


{\bf Acknowledgements}

We thank Simon Johnston for useful discussions regarding the radio polarization signal.

The Nan\c cay Radio Observatory is operated by the Paris Observatory, associated with the French 
Centre National de la Recherche Scientifique (CNRS). This research made use of the WEBDA database, 
operated at the Institute for Astronomy of the University of Vienna.

The \textit{Fermi} LAT Collaboration acknowledges generous ongoing support
from a number of agencies and institutes that have supported both the
development and the operation of the LAT as well as scientific data analysis.
These include the National Aeronautics and Space Administration and the
Department of Energy in the United States, the Commissariat \`a l'Energie Atomique
and the Centre National de la Recherche Scientifique / Institut National de Physique
Nucl\'eaire et de Physique des Particules in France, the Agenzia Spaziale Italiana
and the Istituto Nazionale di Fisica Nucleare in Italy, the Ministry of Education,
Culture, Sports, Science and Technology (MEXT), High Energy Accelerator Research
Organization (KEK) and Japan Aerospace Exploration Agency (JAXA) in Japan, and
the K.~A.~Wallenberg Foundation, the Swedish Research Council and the
Swedish National Space Board in Sweden.

Additional support for science analysis during the operations phase is gratefully
acknowledged from the Istituto Nazionale di Astrofisica in Italy and the Centre National d'\'Etudes Spatiales in France.


\bibliographystyle{aa}
\bibliography{PSR0244,Pulsar_Catalog_ALL_Refs}

\end{document}